# An Integrated Risk Assessment Process of Safety-Related Digital I&C Systems in Nuclear Power Plants

Hongbin Zhang,[a,#] Han Bao,[a]* Tate Shorthill,[b] Edward Quinn,[c]

[a]*Idaho National Laboratory, P.O. Box 1625, MS 3860, Idaho Falls, 83415 ID, United States*

[b]*University of Pittsburgh, 3700 O'Hara Street, Pittsburgh, Pennsylvania 15261*

[c]*Technology Resources, Dana Point, CA*

*han.bao@inl.gov

#Current Address: Terrapower, 15800 Northup Way, Bellevue, WA 98008

# An Integrated Risk Assessment Process of Safety-Related Digital I&C Systems in Nuclear Power Plants

Upgrading the existing analog instrumentation and control (I&C) systems to state-of-the-art digital I&C (DI&C) systems will greatly benefit existing light-water reactors (LWRs). However, the issue of software common cause failure (CCF) remains an obstacle in terms of qualification for digital technologies. Existing analyses of CCFs in I&C systems mainly focus on hardware failures. With the application and upgrading of new DI&C systems, design flaws could cause software CCFs to become a potential threat to plant safety, considering that most redundancy designs use similar digital platforms or software in their operating and application systems. With complex multi-layer redundancy designs to meet the single failure criterion, these I&C safety systems are of particular concern in U.S. Nuclear Regulatory Commission (NRC) licensing procedures. In Fiscal Year 2019, the Risk-Informed Systems Analysis (RISA) Pathway of the U.S. Department of Energy's (DOE's) Light Water Reactor Sustainability (LWRS) Program initiated a project to develop a risk assessment strategy for delivering a strong technical basis to support effective, licensable, and secure DI&C technologies for digital upgrades and designs. An integrated risk assessment for the DI&C (IRADIC) process was proposed for this strategy to identify potential key digital-induced failures, implement reliability analyses of related digital safety I&C systems, and evaluate the unanalyzed sequences introduced by these failures (particularly software CCFs) at the plant level. This paper summarizes these RISA efforts in the risk analysis of safety-related DI&C systems at Idaho National Laboratory.

## I. INTRODUCTION

Digital upgrades and plant modernization efforts offer the foremost path to performance and cost improvements of nuclear power plants (NPPs) [1]. Despite decades of experience with analog systems, the technical challenges associated with their continued use (e.g., signal drift, high maintenance costs, obsolescence, and lack of industrial suppliers) have caused the nuclear industry to move toward digital instrumentation and control (DI&C) in favor of integrated circuitry and the modern microcontroller [3]. Compared with analog systems, DI&C systems offer significant advantages in the areas of monitoring, processing, testing, and maintenance [4] [5]. Notwithstanding the immediate attraction, the nuclear industry has been slow to adopt safety-rated DI&C because each new design must be shown to maintain or improve the status quo by means of a risk assessment [3]. Though many of the concepts for the risk assessment of analog systems carry over, DI&C systems present unique challenges. In 1997, the National Research Council detailed several technical challenges for the implementation of DI&C systems. Those relating specifically to the present work are: (1) the system aspects of digital systems; (2) the potential for software-based common cause failures (CCFs); and (3) the need for a risk assessment method tailored to DI&C systems [3].

The system aspects of DI&C involve issues that extend beyond individual components and even beyond the function of the system itself. The challenge with using these system aspects is discussed in NUREG/CR-6901. Digital systems exhibit two types of interactions—Type 1: the interactions of a DI&C system (and/or its components) with a controlled process (e.g., NPP) and Type 2: the interactions of a DI&C system (and/or its components) with itself and/or other digital systems and components [6]. Kirschenbaum et al. provide a useful summary of these concerns in their own work on the investigation of digital systems [7]. Common or redundant components are often utilized as a backup to ensure system reliability. However, the improper application of redundant features can leave a system vulnerable to CCFs, which arise from the malfunction of two or more components, or functions, due to a single failure source [1] [8]. To make redundancy designs effective, diversity is employed, providing an alternative technology, method, technique, or means to achieve a desired result [9]. The diverse protection helps eliminate the common features necessary for a CCF. Early in 1995, the U.S. Nuclear Regulatory Committee (NRC) probabilistic risk assessment (PRA) policy statement believed the use of risk information in all regulatory activities would promote regulatory stability and efficiency to the extent supported by the state-of-the-art in PRA methods and data [10], while diversity and defense-in-depth (D3) analyses were mainly performed using deterministic approaches.

In Fiscal Year (FY) 2019, the Risk-Informed Systems Analysis (RISA) Pathway of the U.S. Department of Energy's (DOE's) Light Water Reactor Sustainability (LWRS) program initiated a project to develop a risk assessment strategy for delivering a strong technical basis to support effective, licensable, and secure DI&C technologies for digital upgrades/designs [11] [12] [13]. An integrated risk assessment for the DI&C (IRADIC) process was proposed for this strategy, which aims to identify key digital-induced failures, implement reliability analyses on related digital safety I&C systems, and evaluate the unanalyzed sequences introduced by these failures (particularly software CCFs) at the plant level. More details are included in Section II. According to the guidelines and requirements of the IRADIC process, an approach for redundancy-guided systems-theoretic hazard analysis (RESHA) was developed in FY-2020. It aims to help system designers and engineers identify digital-based CCFs and qualitatively analyse their effects on digital system vulnerability. It also provides a technical basis for implementing future reliability and consequence analyses of unanalyzed sequences and optimizing the D3 applications in a cost-effective way. This approach has been developed and applied for the hazard analysis of digital reactor trip system (RTS) and engineered safety features actuation system (ESFAS). Relevant description and case studies are shown in Section III. A method for software reliability assessment of digital control systems with consideration for the quantification of CCFs is described in Section IV, which is defined as a Bayesian and HRA (human reliability analysis)-aided method for the reliability analysis of software (BAHAMAS). Section V describes the efforts in consequence analysis that evaluate the impact of digital-based failures to the plant safety. Section VI summarizes the conclusion and future work on risk assessment of DI&C systems.

**II. INTEGRATED RISK ASSESSMENT PROCESS FOR DI&C SYSTEMS**

The overall goal of developing an integrated risk assessment approach is to deliver a strong technical basis to support effective, licensable, and secure technologies for DI&C upgrades/designs. To deal with the expensive licensing justifications from regulatory insights, this technical basis is instructive for nuclear vendors and utilities to effectively lower the costs associated with digital compliance and speed industry advances by: (1) defining an integrated risk-informed analysis process for DI&C upgrade, including hazard analysis, reliability analysis, and consequence analysis; (2) applying systematic and risk-informed tools to identify CCFs and quantify responding failure probabilities for DI&C technologies; (3) evaluating the impact of digital failures at the component level, system level, and plant level; (4) providing insights and suggestions on designs to manage the risks, thereby supporting the development, licensing, and deployment of advanced DI&C technologies on NPPs.

It is critical for the viability of a nuclear power fleet to upgrade DI&C (i.e., safety and non-safety-related) systems in existing NPPs within a cost-effective and regulatory acceptable way. One key outcome of this project is to perform a plant-specific risk assessment to provide sustainable scientific support for enabling industry to balance the digital-related risk and cost.

The IRADIC technology consists of two parts: risk analysis and risk evaluation. Risk analysis, including hazard analysis, reliability analysis, and consequence analysis, focuses on identifying potential failures of digital systems and components, estimating probabilities, and analyzing relevant consequences. Risk evaluation compares risk analysis results with specific risk acceptance criteria in component, system, and plant levels. Figure 1 displays the schematic of the IRADIC technology for safety evaluation and design optimization of DI&C systems. More details about the workflows and information flows of hazard, reliability and consequence analysis can be found in Sections III, IV, and V. The IRADIC technology was also suggested to deal with the software risk analysis for digital twins in the nearly autonomous management and control systems [14].

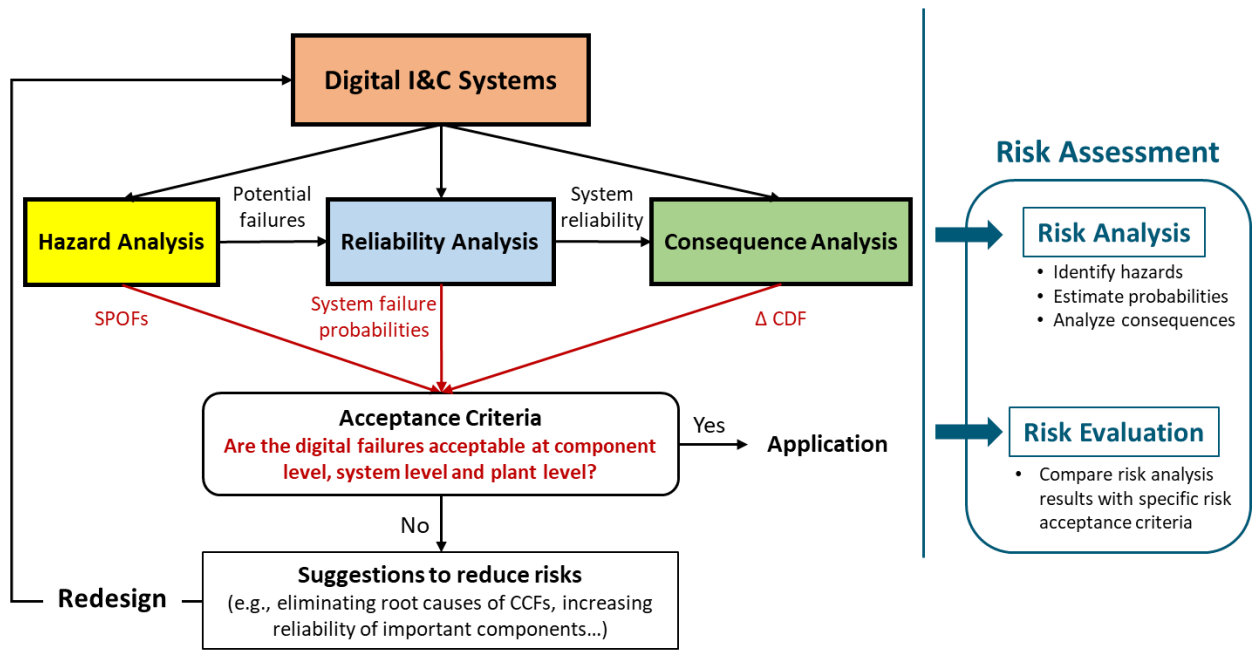

Figure 1. Schematic of the IRADIC technology for safety evaluation and design optimization of DI&C systems.

## III. REDUNDANCY-GUIDED SYSTEM-THEORETIC HAZARD ANALYSIS

In the IRADIC framework, a method for hazard analysis, RESHA [15][16], was developed by combining fault tree analysis (FTA) along with a reframed, redundancy-guided application of the systems-theoretic process analysis (STPA) [17]. An integrated fault tree (FT) can be generated with both software failures and hardware failures and used to discover single points of failure (SPOFs) leading to the loss of function of the entire DI&C system. SPOF refers to a situation in which a single part of a system fails, and the entire system loses function as a result. The proposed approach for a RESHA is illustrated in Figure 2. RESHA, its steps, and the role of STPA within this hazard analysis are briefly described in the subsequent paragraphs.

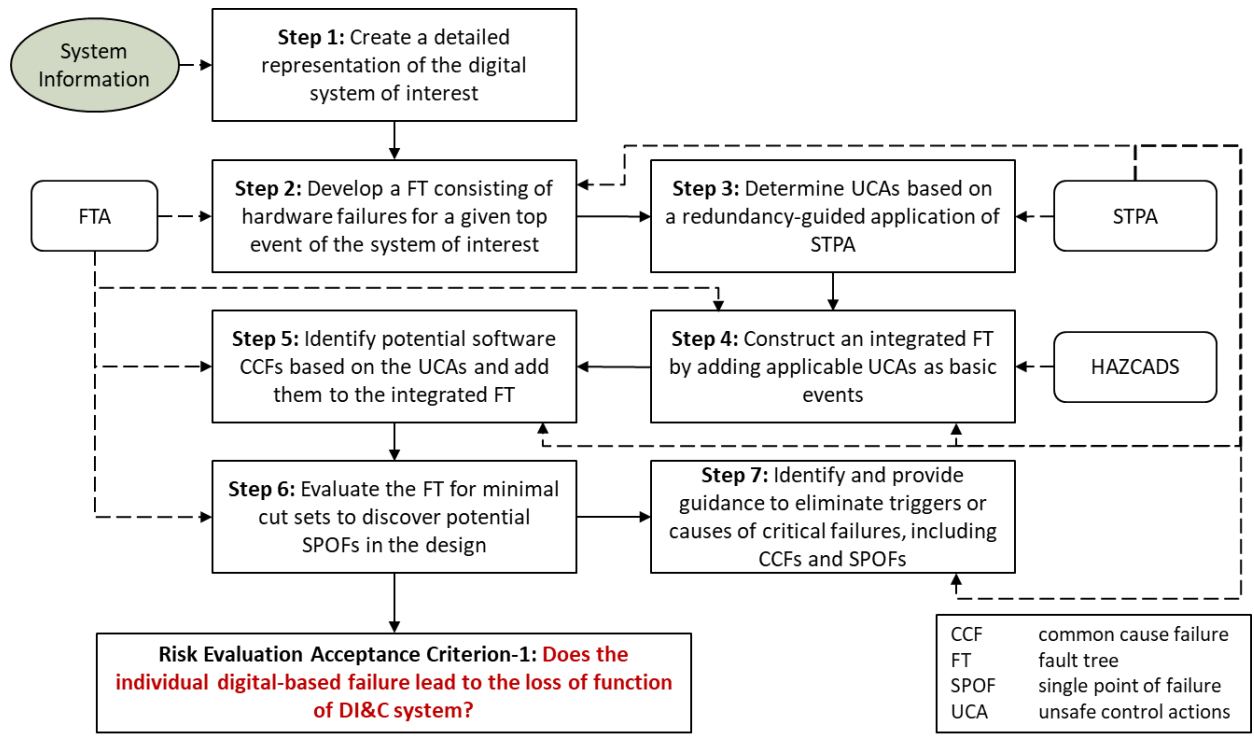

Figure 2. Workflow of the proposed RESHA approach (derived from [15] and [16]). The dashed lines indicate the influence of FTA, STPA [17], and HAZCADS [18].

RESHA employs a reframed STPA to identify software-based failure events to be used within an integrated FT. STPA was designed as a systems-focused, top-down approach to modeling [17]. The goal of STPA is to assess a system for unsafe behavior and identify scenarios that link that behavior to system-level hazards and losses. The system-level hazards are those system states or conditions that may lead to a loss (i.e., something of value to stakeholders) [17]. STPA consists of four main parts. The first three parts emphasize the identification of undesirable or unsafe control actions (UCAs), and the last part determines the context or scenario for which a UCA might occur. There are four categories of UCAs in STPA: (1) control action is not provided when it is needed; (2) control action is provided when it is not needed; (3) control action is provided when it is needed but too early, too late, or in a wrong order; (4) control action lasts too long or stops too soon (only applicable to continuous control actions) [17]. In a digital system, unsafe or undesirable control actions and information exchanges may lead to a failure of the digital system; hence, UCAs and their causes are selected as potential software failures. RESHA relies on STPA concepts to support a redundancy-guided hazard analysis.

The first step of RESHA, like most methods for hazard analysis, is focused on information gathering; this might require the creation of diagrams and sketches. The essential aspect of this step is to assemble the necessary information for the remaining steps. Step 2 begins the formation of a FT that serves as the backbone for risk assessment within IRADIC. In this step, the structure of the FT is created based on the hardware components of the system of interest. While not required, the FT is often linked to an event tree (ET) as part of an event tree analysis (ETA); that link is the FT's top-event [1]. Proper selection of the top event is a vital aspect of both FTA and ETA. Here, STPA can be leveraged to inform decisions regarding the selection of FT top events and their relationships to an ET. The STPA-identified system-level hazards and losses can provide clarification for what the RESHA integrated FT will look like; system-level hazards may serve as FT top events that integrate with an ET for tracking system-level losses. FT top events aid in selecting credible UCA from those identified in Step 3.

The goal of Step 3 is to identify UCAs as potential software failures by means of a redundancy-guided application of STPA. UCAs, and their causes, are selected as the potential software failures to be included within FT from Step 2. In order to find UCAs, STPA relies on a control structure that details the controllers and processors of the system. STPA does not explicitly model safety features such as redundant components in the control structure; these details are left to be addressed in context scenario discussions during the final part of STPA [17]. Failure to provide explicit incorporation of the safety features early and within the control structure diagram may cause the potential CCFs in redundant designs to be overlooked. Thus, STPA is reframed according to safety features (e.g., redundant and diverse designs) directly and early by explicitly modeling them within the control structure diagrams.

A redundancy-guided multi-layer control structure is formed by decomposing the system based on redundancy. A top-down approach identifies functional redundancy within the system and creates a control structure for the components and modules pertaining to that layer. The process is repeated systematically and incorporating the information exchanges found for each component within each redundancy layer. The result is a multi-layer control structure that captures all the necessary control actions of the system and its components. Finally, UCAs are identified from the control signals indicated in the control structure diagram.

Step 4 combines the FT with UCAs from STPA, a concept borrowed from the hazard and consequence analysis for digital systems (HAZCADS) [18]. In this step, applicable UCAs are selected and added into the hardware FT as the software failures. For a specific top event in the FT, some UCAs may be inapplicable. For example, the UCA of a component associated with UCA category 1 (i.e., "action not provided") may not be applicable for a top event that represents spurious activation. Applicable UCAs are added to the FT as potential software failure events. Step 5 provides consideration of CCFs. Any group of identical components may be susceptible to a CCF that falls into any of the categories of UCAs. These groups of components are called common cause component groups (CCCGs). Basic events that represent CCFs are also added based on applicable UCA categories. In some instances, the redundancy layers of the multi-layered control structure may distinguish CCCGs and provide indication for which CCF basic events should be added. For example, the CCF of a CCCG associated with a particular redundant division from a four-division digital system. After adding UCAs and CCFs to the FT, the next step performs a qualitative evaluation of the integrated FT.

Step 6 provides the main outcome of the hazard analysis; the minimal cut sets of the integrated FT are evaluated to determine potential critical points of failure. The critical points of failure are the low-order cut sets (i.e., those cut sets with few basic events). Of particular interest are the SPOFs. Identification of the basic events that make up the low-order cut sets provides a starting place for potential design improvements and for reliability analysis.

The purpose of Step 7 is to identify and provide guidance to eliminate latent faults or triggers of CCFs and other critical failures or UCAs identified in Step 6. The STPA Handbook [17] indicates that the causes of UCAs can be grouped into two categories: (1) unsafe controller behaviors and (2) inadequate feedback and/or other inputs. STPA provides guidance for identification of these categories based on how controllers process information and act.

Currently, RESHA has been demonstrated for the hazard analysis of a four-division digital RTS [15] and ESFAS [16]. The designs for the ESFAS and RTS in those works have similar structures to state-of-the-art digital systems in existing NPP designs such as the APR-1400 [19]. Portions of FT for RTS failure with software failures are displayed in Figure 3, Figure 4, Figure 5, and Figure 6. More details can be found in [15].

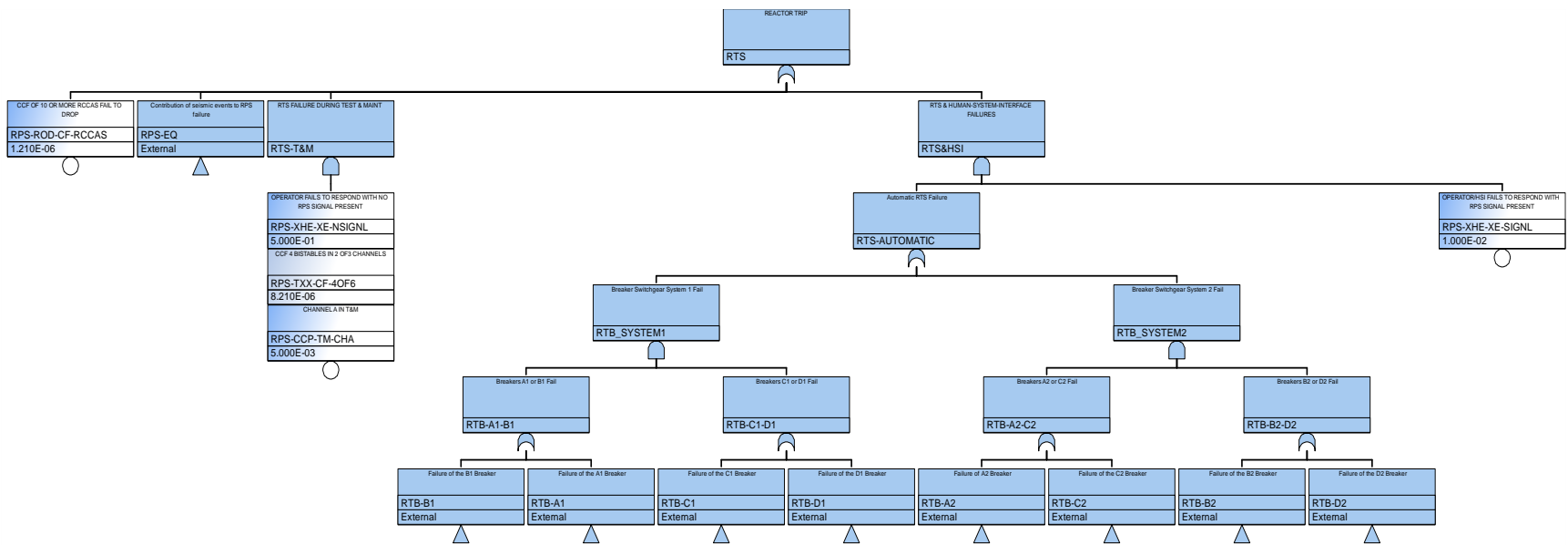

Figure 3. Main FT of integrated RTS-FT using IRADIC technology.

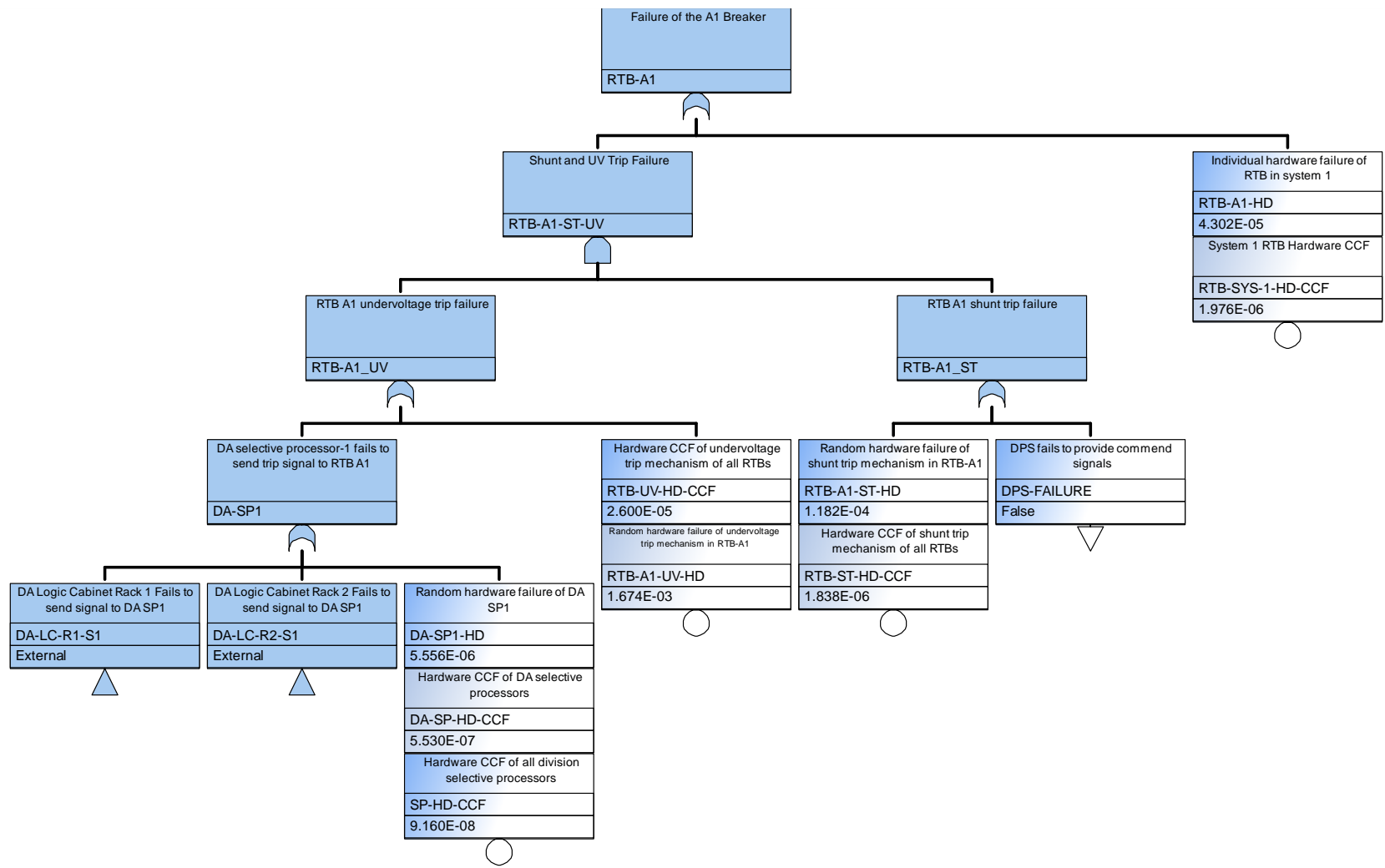

Figure 4. Transfer event of "Failure of A1 Breaker" of the integrated RTS-FT.

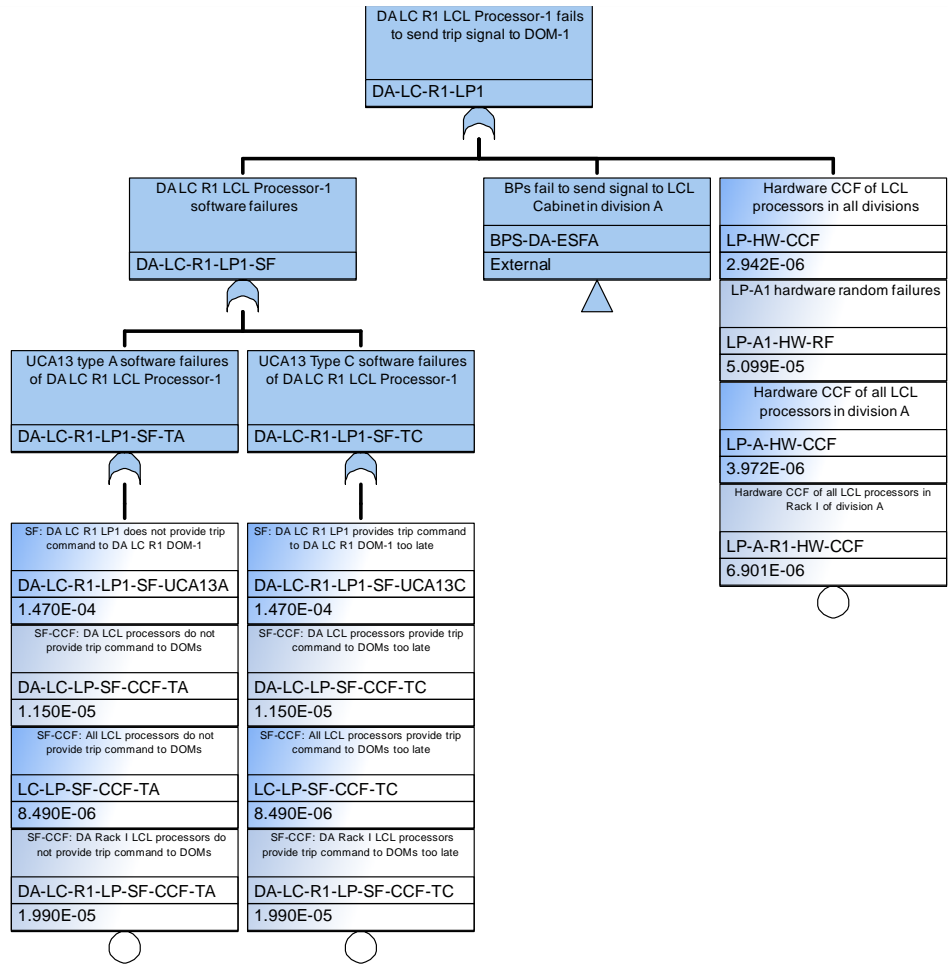

Figure 5. Transfer event of "DA LC R1 LCL Processor-1 fails to send trip signal to DOM-1" of the integrated RTS-FT.

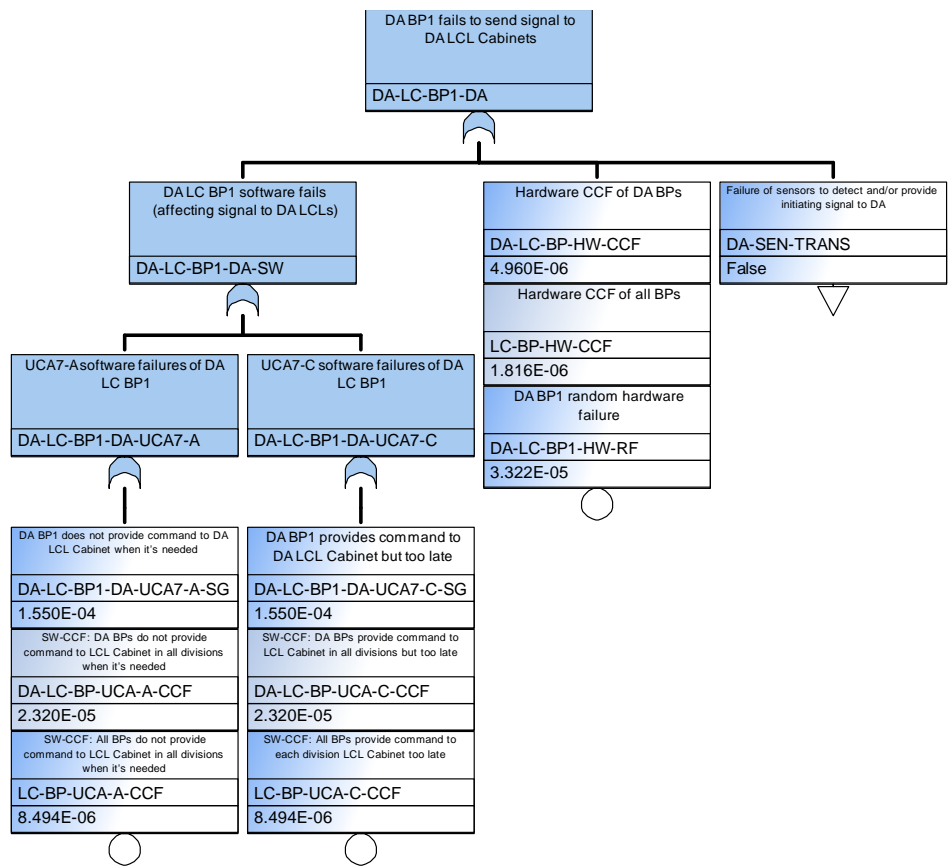

Figure 6. Transfer event of "DA BP1 fails to send signal to DA LCL Cabinets" of integrated RTS-FT.

## IV. INTEGRATED RELIABILITY ANALYSIS

The reliability analysis in IRADIC consists of (1) quantification of the basic events of the integrated FT that is built up using RESHA and (2) estimation of top event of the integrated FT. In this work, the BAHAMAS method is applied to quantify software failure probability; hardware failure probabilities are collected from previous publications [20]. The quantification of integrated FT is performed using the Idaho National Laboratory (INL)-developed PRA tool Systems Analysis Programs for Hands-on Integrated Reliability Evaluations (SAPHIRE) [21].

### IV.A. Reliability Analysis of Software of a Four-division Digital Reactor Trip System

BAHAMAS was developed by INL for reliability estimations of software in early development stage [22]. It can provide a rough estimation of software failure probabilities even when testing/performance data of the target software is very limited. In this condition, BAHAMAS assumes that software failures root from human errors in the software development life cycle (SDLC) and can be modeled and roughly estimated using HRA. In BAHAMAS, a Bayesian belief network (BBN) is constructed to integrate disparate causal factors of the system in a logic way. More technical information about BAHAMAS method can be found in [22].

The BAHAMAS workflow is briefly introduced in this section, where each of the main methods mentioned in the approach is incorporated for the reliability analysis of a software system. As discussed in Section I, the risk assessment of digital systems has been divided into three phases. Phase 2 provides quantification for the results found in Phase 1. Although it is the intention in Phase 2 for BAHAMAS to be flexible, much of its formulation is based on the results of a RESHA-based Phase 1 hazard analysis. Consequently, the subsequent approach to Phase 2 is tailored best to hazards identified by RESHA. BAHAMAS workflow and information flow are shown in Figure 7 and Figure 8, respectively.

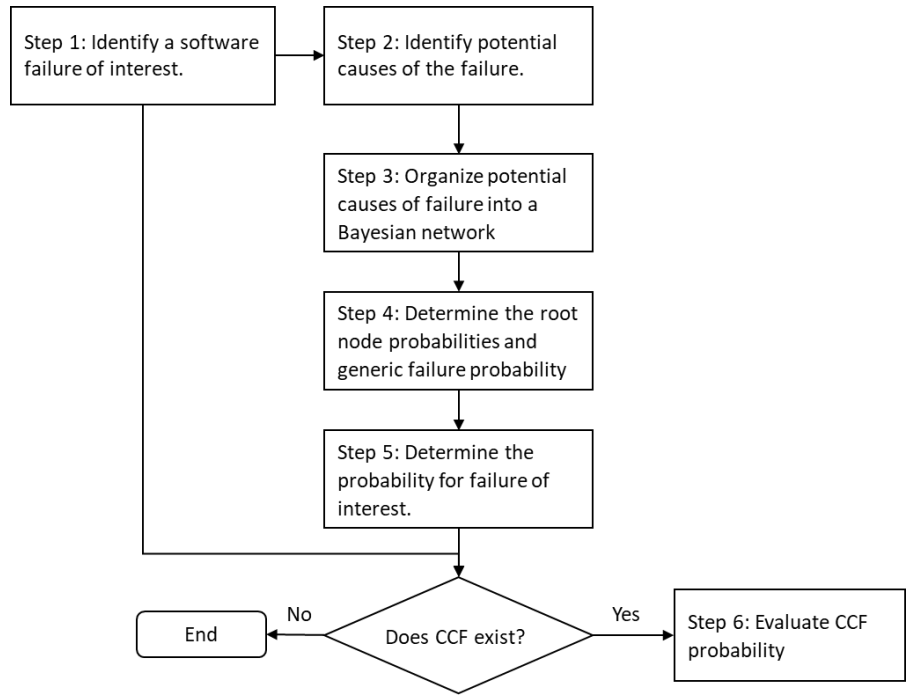

Figure 7. BAHAMAS workflow.

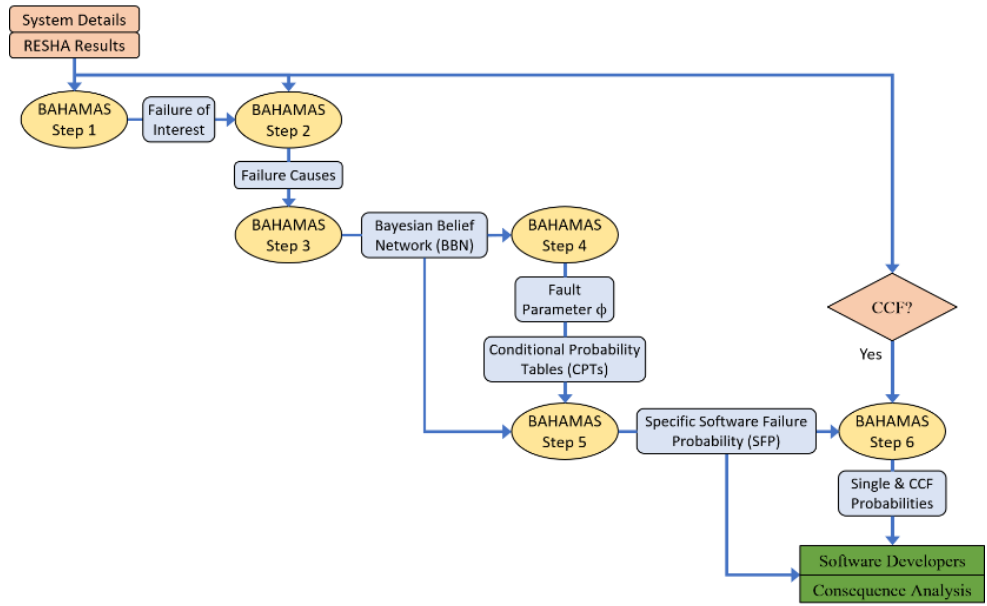

Figure 8. Flowchart showing the primary inputs and outputs of each step of BAHAMAS (derived from [22]).

Step 1 of BAHAMAS is to select a software failure of interest that needs to be quantified from a qualitative study (e.g., RESHA). For example, it can be an individual failure or CCF of bistable processor of the four-division RTS that was identified in RESHA study. Step 2 collects information regarding the event of interest and identifies potential causes of the failure of interest. For the UCA of bistable processors, the root causes may be software inner defects or data communication errors due to environmental hazards or human errors. Step 3 builds up a BBN to organize the potential causes that were identified in Step 2 so that an acyclic (i.e., without feedback) graphical network representing the relationships of interest can be used for quantification process. In this case, this refers to the relationships between root causes and probability of software failure. Step 4 determines the fault parameter by estimating the root node probabilities and generic software failure probability. Step 5 determines the probability for failure of interest by estimating specific software failure probability and evaluating single and CCF probability. Step 6 conducts CCF modeling and estimation using the beta-factor method [23], which assumes the total failure probability of a component is the sum of the individual and the CCF probabilities. A beta-factor is estimated based on existing data to represent the proportions of the individual failures and the CCFs in the total failure probability. A case study has been performed in [22]; for the software failure of bistable processors of the four-division RTS, the individual failure probability is 1.554E-4, the probability of the CCF of bistable processor in all divisions is 8.494E-6, and the probability of the CCF of bistable processor in one division is 2.320E-5.

**IV.B. Quantification of the Integrated Fault Tree of a Four-division Digital Reactor Trip System**

By assigning the software and hardware failure probabilities into the integrated FTs of the four-division digital RTS, the failure probabilities of the RTS can be calculated using INL-developed PRA tool SAPHIRE. The RTS failure probability is 1.270E-6. Mechanical CCF of rod control cluster assembly (RCCA) is the main contributor to the failure of the representative four-division digital RTS; the software CCFs do not have significant impacts to the failure of digital RTS because of the highly redundant design and high reliability of digital components.

## V. CONSEQUENCE ANALYSIS

This section describes the consequence analysis of a generic pressurized-water reactor (PWR) SAPHIRE model with integrated FT for a four-division digital RTS. This model was developed using SAPHIRE 8 for a typical PWR plant for the accident scenario analysis with an original FT for a two-division analog RTS. The core damage frequency (CDF) has been calculated for the ET model with different RTS-FTs and compared to show how much safety margin can be increased by introducing the modern four-division digital RTS.

The original two-train analog RTS was modeled with different failure modes, such as electric failures, CCF of RCCA fail to drop, contribution of seismic events, operator errors, and RTS failures during test and maintenance. The FT was quantified using SAPHIRE 8, and the RTS failure probability is 4.288E-6. It shows that the failure probability of integrated four-division digital RTS-FT is only about 50% of the original one.

In this paper, an accident scenario about INT-TRANS (initiating event - general plant transient) is selected for the consequence analysis of a four-division digital RTS failure, the ET model is shown in Figure 9 and Figure 10.Table 1 compares the values of CDF with original and new RTS-FTs. The original total IE-TRANS CDF is 1.073E-6/reactor year and greatly reduced to 6.418E-07/reactor year with the new RTS-FTs. There are 16 non-zero CDF sequences out of a total of 145 INT-TRANS accident sequences (i.e., the sequence end state is core damage).

INT-TRANS:21-16 from ATWS (anticipated transient without scram) scenarios is one of the most risk-significant sequences with a CDF reduced from 5.388E-7/reactor year to 1.596E-7/reactor year and contributes 24.87% of the CDF of improved INT-TRANS. In this sequence, RTS fails to trip the reactor; primary and secondary side depressurizations are not successful because safety relief values are closed. Core damage occurs as long-term cooling cannot be established.

INT-TRANS:21-14 from ATWS scenarios is another risk-significant sequence with a CDF reduced from 7.262-8/reactor year to 2.150E-08/reactor year, contributing 3.35% of the CDF of improved INT-TRANS. In this sequence, RTS fails to trip the reactor; reactor cooling system fails to limit the pressure under 3200 psi; main feedwater is unavailable and emergency boration fails. Core damage occurs as long-term cooling cannot be established.

Results show that by introducing a four-division digital RTS instead of the two-division analog RTS, the safety margin increased from the plant digitalization on safety-related DI&C system can be quantitatively estimated: the CDF is significantly reduced. Plant modernization including the improvement of safety-related DI&C systems such as RTS will benefit plant safety by providing more safety margins.

In addition, the number of cut sets is also reduced from 3590 to 3474 due to the improved design from a two-train analog system to a four-division digital system. As the complexity of the system increases, the number of failure combination should also increase. However, with the improved design, the cut-set probabilities are reduced and truncated below the 1E-12 threshold.

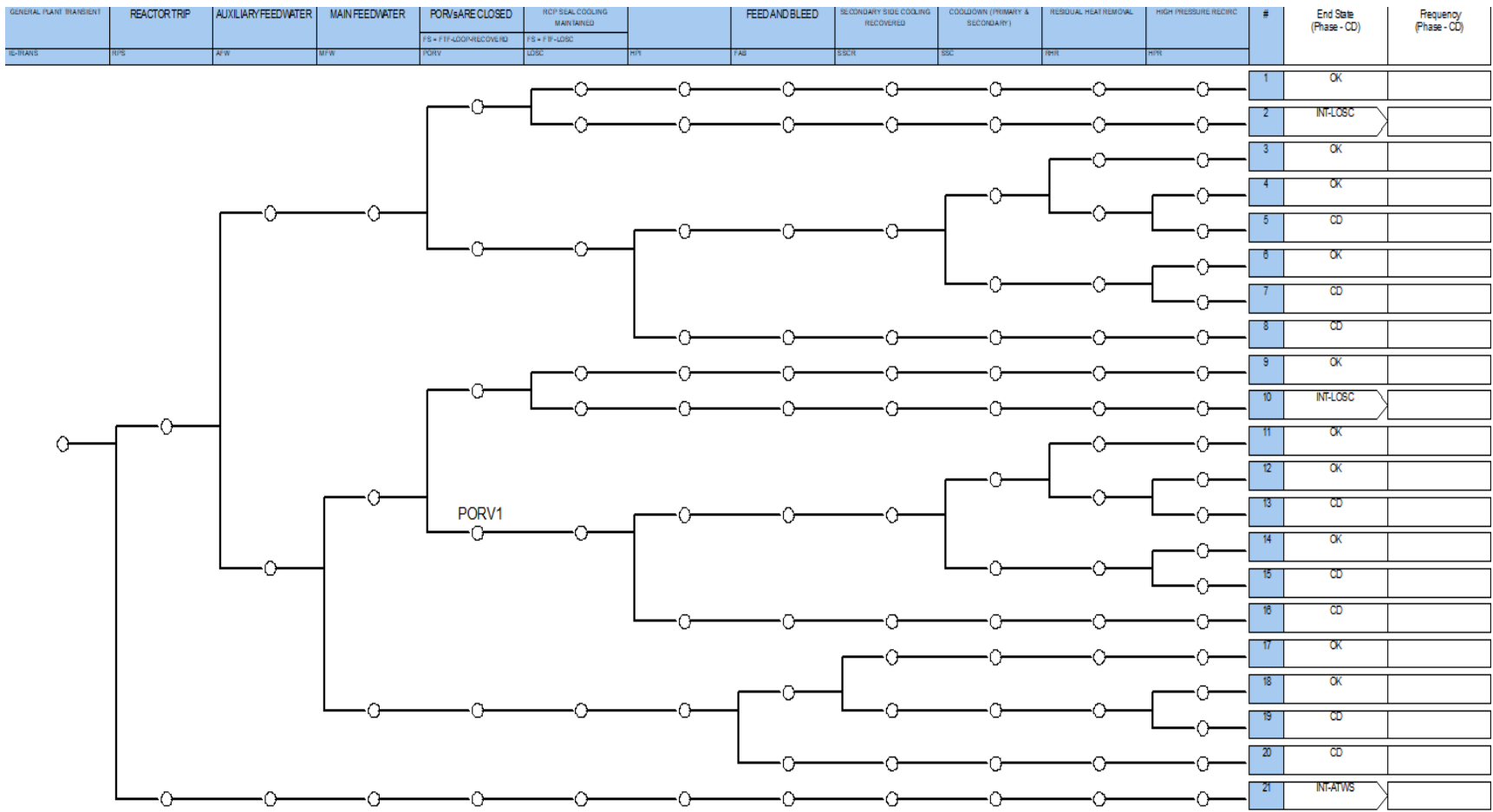

Figure 9. Generic PWR ET for general plant transient (INT-TRANS).

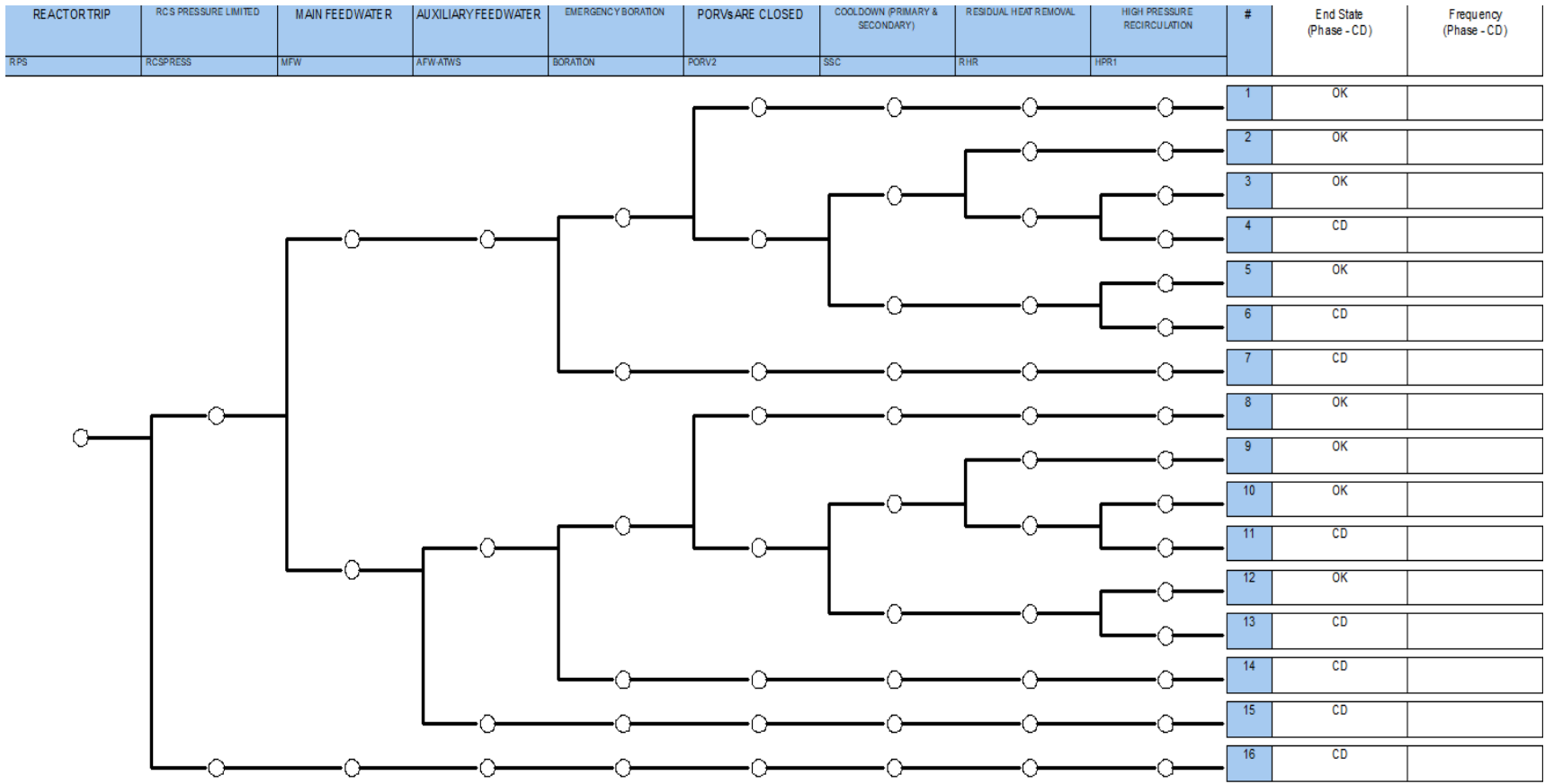

Figure 10. Generic PWR ET for anticipated transient without scram scenarios (INT-ATWS).

Table 1. Comparison of INT-TRANS ET quantification results.

| Sequence | CDF (per reactor year) | | | # of Cut Sets (Truncated by 1E-12) | |
| --- | --- | --- | --- | --- | --- |
| | Original ET | Improved ET | Δ CDF/ Original CDF | Original ET | Improved ET |
| INT-TRANS:21-16 | 5.388E-07 | 1.596E-07 | -70.38% | 51 | 38 |
| INT-TRANS:20 | 3.895E-07 | 3.895E-07 | 0 | 1060 | 1060 |
| INT-TRANS:21-14 | 7.262E-08 | 2.150E-08 | -70.40% | 49 | 18 |
| INT-TRANS:02-02-09 | 5.830E-08 | 5.830E-08 | 0 | 1248 | 1248 |
| INT-TRANS:19 | 8.132E-09 | 8.132E-09 | 0 | 282 | 282 |
| INT-TRANS:02-03-09 | 2.731E-09 | 2.731E-09 | 0 | 387 | 387 |
| INT-TRANS:02-02-10 | 9.546E-10 | 9.546E-10 | 0 | 168 | 168 |
| INT-TRANS:21-15 | 7.568E-10 | 2.149E-10 | -70.60% | 102 | 30 |
| INT-TRANS:02-04-10 | 5.865E-10 | 5.865E-10 | 0 | 142 | 142 |
| INT-TRANS:02-14-10 | 1.994E-10 | 1.994E-10 | 0 | 81 | 81 |
| INT-TRANS:02-03-10 | 7.653E-12 | 7.653E-12 | 0 | 4 | 4 |
| INT-TRANS:02-09-09 | 7.558E-12 | 7.558E-12 | 0 | 4 | 4 |
| INT-TRANS:02-06-09 | 7.558E-12 | 7.558E-12 | 0 | 4 | 4 |
| INT-TRANS:02-08-09 | 7.558E-12 | 7.558E-12 | 0 | 4 | 4 |
| INT-TRANS:02-07-09 | 2.287E-12 | 2.287E-12 | 0 | 2 | 2 |
| INT-TRANS:02-10-09 | 2.287E-12 | 2.287E-12 | 0 | 2 | 2 |
| Total | 1.073E-06 | 6.418E-07 | -40.19% | 3590 | 3474 |

## VI. CONCLUSIONS AND FUTURE WORK

This paper summarized the development of an integrated risk assessment technology for highly redundant safety-related DI&C systems in NPPs. By integrating hazard analysis, reliability analysis, and consequence analysis together, the risk assessment strategy aims to: (1) help system designers and engineers to systematically address digital-based CCFs and quantitatively analyze their effects on digital system vulnerability and key plant responses; (2) improve existing PRA models for the industry by identifying and evaluating the risk associated with DI&C technologies; and (3) provide risk insights to address the licensing challenges facing DI&C upgrades. Results show that by adding the integrated FT of the four-division digital RTS instead of the two-division analog RTS, the safety margin increased from the plant digitalization on a safety-related DI&C system can be quantitatively estimated; the CDF is significantly reduced. It indicates that plant modernization including the improvement of safety-related DI&C systems such as RTS will benefit plant safety by providing more safety margins.

One area for future research of LWRS-RISA is to deal with the risk analysis for the Human System Interface (HSI) in DI&C modernization of existing NPPs. The HSI is one of the key advanced design features applied in modern DI&C systems of NPPs. Normally, it is designed based on a compact workstation-based system in the control room. The compact workstation provides a convenient operating environment to facilitate the display of plant status information to the operator so that operability is enhanced by using advanced display, alarm, and procedure systems. The HSI should have sufficient diversity to demonstrate D3 protection against CCF of the safety system. However, the vulnerability of HSI is affected by many factors, such as human errors, cyber-attacks, software CCFs, etc. Therefore, one of the future works aims to identify, evaluate, and reduce these system vulnerabilities to support the licensing, deployment, and operation of the HSI designs. Relevant research results will be published soon. Another future work is uncertainty quantification and verification of RESHA and BAHAMAS.

## VII. ACKNOWLEDGMENTS

This submitted manuscript was authored by a contractor of the U.S. Government under DOE Contract No. DE-AC07-05ID14517. Accordingly, the U.S. Government retains and the publisher, by accepting the article for publication, acknowledges that the U.S. Government retains a nonexclusive, paid-up, irrevocable, worldwide license to publish or reproduce the published form of this manuscript, or allow others to do so, for U.S. Government purposes. This information was prepared as an account of work sponsored by an agency of the U.S. Government. Neither the U.S. Government nor any agency thereof, nor any of their employees, makes any warranty, express or implied, or assumes any legal liability or responsibility for the accuracy, completeness, or usefulness of any information, apparatus, product, or process disclosed, or represents that its use would not infringe privately owned rights. References herein to any specific commercial product, process, or service by trade name, trademark, manufacturer, or otherwise, does not necessarily constitute or imply its endorsement, recommendation, or favoring by the U.S. Government or any agency thereof. The views and opinions of authors expressed herein do not necessarily state or reflect those of the U.S. Government or any agency thereof.

## VIII. REFERENCES

1. M. STAMATELATOS, W. VESELY, J. DUGAN, J. FRAGOLA, J. MINARICK III and J. RAILSBACK, "Fault Tree Handbook with Aerospace Applications," Version 1.1, National Aeronautics and Space Administration, Washington, DC (2002).
2. K. THOMAS AND K. SCAROLA, "Strategy for Implementation of Safety-Related Digital I&C Systems," INL/EXT-18-45683, Idaho National Laboratory (June 2018).
3. National Research Council, *Digital Instrumentation and Control Systems in Nuclear Power Plants: Safety and Reliability Issues*, Washington, DC: The National Academies Press (1997).
4. H. HASHEMIAN, "Nuclear Power Plant Instrumentation and Control," in Nuclear Power - Control, Reliability and Human Factors, P. Tsvetkov, Ed., pp. 49-66, Intech (2011); https://doi.org/10.5772/18768.
5. T.-L. CHU, M. YUE, G. MARTINEZ-GURIDI, and J. LEHNER, "Review of Quantitative Software Reliability Methods," BNL-94047-2010, Brookhaven National Laboratory (September 2010); https://doi.org/10.2172/1013511.
6. T. ALDEMIR, D. MILLER, M. STOVSKY, J. KIRSCHENBAUM, P. BUCCI, A. FENTIMAN, and L. MANGAN, "Current State of Reliability Modeling Methodologies for Digital Systems and Their Acceptance Criteria for Nuclear Power Plant Assessments," NUREG/CR-6901, U.S. Nuclear Regulatory Commission (February 2006).

7. J. KIRSCHENBAUM, P. BUCCI, M. STOVSKY, D. MANDELLI, T. ALDEMIR, M. YAU, S. GUARRO, E. EKICI, and S. A. ARNDT, "A Benchmark System for Comparing Reliability Modeling Approaches for Digital Instrumentation and Control Systems," *Nuclear Technology*, **165**, *1*, 53 (2009); https://doi.org/10.13182/NT09-A4062.

8. T. E. WIERMAN, D. M. RASMUSON, and A. MOSLEH, "Common-Cause Failure Databased and Analysis System: Event Data Collection, Classification, and Coding," NUREG/CR-6268, Rev. 1, Idaho National Laboratory, (September 2007).

9. U.S. Nuclear Regulatory Commission, *A Defense-In-Depth and Diversity Assessment of the RESAR-414 Integrated Protection System*, U.S. Nuclear Regulatory Commission, Washington, DC (1979).

10. U.S. NRC, " Use of Probabilistic Risk Assessment Methods in Nuclear Regulatory Activities," 95-20237, U.S. NRC, (1995). https://www.federalregister.gov/documents/1995/08/16/95-20237/use-of-probabilistic-risk-assessment-methods-in-nuclear-regulatory-activities-final-policy-statement.

11. H. BAO, H. ZHANG, and K. THOMAS, "An Integrated Risk Assessment Process for Digital Instrumentation and Control Upgrades of Nuclear Power Plants," INL/EXT-19-55219, Idaho National Laboratory, (August 2019).

12. H. BAO, T. SHORTHILL, and H. ZHANG, "Redundancy-guided System-theoretic Hazard and Reliability Analysis of Safety-related Digital Instrumentation and Control Systems in Nuclear Power Plants," INL/EXT-20-59550, Idaho National Laboratory (August 2020).

13. H. BAO, T. SHORTHILL, E. CHEN, and H. ZHANG, "Quantitative Risk Analysis of High Safety-significant Safety-related Digital Instrumentation and Control Systems in Nuclear Power Plants using IRADIC Technology," INL/EXT-21-64039, Idaho National Laboratory (August 2021).

14. L. LIN, H. BAO, and N. DINH, "Uncertainty quantification and software risk analysis for digital twins in the nearly autonomous management and control systems: A review," *Annals of Nuclear Energy*, **160**, 108362 (2021); https://doi.org/10.1016/j.anucene.2021.108362.

15. T. SHORTHILL, H. BAO, H. ZHANG, and H. BAN, "A Redundancy-Guided Approach for the Hazard Analysis of Digital Instrumentation and Control Systems in Advanced
25

Nuclear Power Plants," *Nuclear Technology* (2021); https://doi.org/10.1080/00295450.2021.1957659.

16. H. BAO, T. SHORTHILL, and H. ZHANG, "Hazard Analysis for Identifying Common Cause Failures of Digital Safety Systems using a Redundancy-Guided Systems-Theoretic Approach," *Annals of Nuclear Energy*, **148**, 107686 (2020).
17. N. G. LEVESON and J. P. THOMAS, *STPA Handbook* (2018).
18. J. CLARK, A. D. WILLIAMS, A. MUNA, and M. GIBSON, "Hazard and Consequence Analysis for Digital Systems – A New Approach to Risk Analysis in the Digital Era for Nuclear Power Plants," *Transactions of the American Nuclear Society*, **119**, *1*, 888, (2018).
19. "APR1400 Design Control Document Tier 2. Chapter 7: Instrumentation and Controls," Korea Electric Power Corporation, Korea Hydro & Nuclear Power Co., Ltd, Korea, Republic of (2018).
20. P. V. VARDE, J. G. CHOI, D. Y. LEE, and J. B. HAN, "Reliability analysis of protection system of advanced pressurized water reactor - APR 1400," KAERI/TR--2468/2003. Korea Atomic Energy Research Institute, Taejon, Korea, Republic of (2003).
21. U.S. Nuclear Regulatory Commission, "Systems Analysis Programs for Hands-on Integrated Reliability Evaluations (SAPHIRE) Version 8.0," NUREG/CR-7039, U.S. Nuclear Regulatory Commission (June 2011).
22. T. SHORTHILL, H. BAO, Z. HONGBIN, and H. BAN, "A novel approach for software reliability analysis of digital instrumentation and control systems in nuclear power plants," *Annals of Nuclear Energy*, **158**, 108260 (2021); https://doi.org/10.1016/j.anucene.2021.108260.
23. D. KANCEV and M. CEPIN, "A new method for explicitly modelling of single failure event within different common cause failure groups," *Reliability Engineering and System Safety*, **103**, 84-93, (2012).